 \documentclass[reprint,prl,superscriptaddress,floatfix]{revtex4-1}

\usepackage{amsmath,amssymb}
\usepackage{graphicx}
\usepackage{color}
\usepackage{soul}
\usepackage{ulem}
\usepackage{multirow}
\usepackage{lineno}
\usepackage{ulem}

\allowdisplaybreaks

\newcommand{\avg}[1]{\left\langle #1 \right\rangle}
\newcommand{\kT}{k_\text{B}T}
\newcommand{\omb}{\omega_\text{b}}

\newsavebox\CBox
\newcommand\hcancel[2][0.5pt]{{\color{red}{%
  \ifmmode\sbox\CBox{$#2$}\else\sbox\CBox{#2}\fi%
  \makebox[0pt][l]{\usebox\CBox}%  
  \rule[0.5\ht\CBox-#1/2]{\wd\CBox}{#1}}}}

\begin{document}

\title{Transition State Theory for solvated reactions beyond recrossing-free dividing surfaces}
\author{F. Revuelta}
\affiliation{Grupo de Sistemas Complejos,
Escuela T\'ecnica Superior de Ingenier\'ia Agron\'omica, 
Alimentaria y de Biosistemas,
Universidad Polit\'ecnica de Madrid,
Avda.~Complutense s/n 28040 Madrid, Spain.}
\affiliation{Instituto de Ciencias Matem\'aticas (ICMAT), 
Cantoblanco, 28049  Madrid, Spain.}
\author{Thomas Bartsch}
\affiliation{Department of Mathematical Sciences,
Loughborough University,
Loughborough LE11 3TU, United Kingdom.}
\author{P. L. Garcia--Muller}
\affiliation{Centro de Investigaciones Energ\'eticas 
Medioambientales y Tecnol\'ogicas, 
Avda.~Complutense 40, 28040 Madrid, Spain.}
\author{Rigoberto Hernandez}
\affiliation{Center for Computational Molecular Sciences and Technology,
School of Chemistry and Biochemistry,
Georgia Institute of Technology,
Atlanta, Georgia 30332-0430, USA}
\author{R. M. Benito}
\affiliation{Grupo de Sistemas Complejos,
Escuela T\'ecnica Superior de Ingenier\'ia Agron\'omica, 
Alimentaria y de Biosistemas,
Universidad Polit\'ecnica de Madrid,
Avda.~Complutense s/n 28040 Madrid, Spain.}
\author{F. Borondo}
\affiliation{Instituto de Ciencias Matem\'aticas (ICMAT), 
Cantoblanco, 28049  Madrid, Spain.}
\affiliation{Departamento de Qu\'imica, 
Universidad Aut\'onoma de Madrid, Cantoblanco, 
28049  Madrid, Spain.}

\date{\today}

\begin{abstract}
The accuracy of rate constants calculated using transition 
state theory depends crucially on the correct 
identification of a recrossing--free dividing surface.
We show here that it is possible to define such optimal 
dividing surface in systems with non--Markovian friction.
However, a more direct approach to rate calculation is based on invariant manifolds
and avoids the use of a dividing surface altogether,
Using that method we obtain an explicit expression for the rate of
crossing an anharmonic potential barrier.
The excellent performance of our method is 
illustrated with an application to a realistic model 
for LiNC$\rightleftharpoons$LiCN isomerization.
\end{abstract}

\pacs{82.20.Db, 05.40.Ca, 05.45.2a, 34.10.+x}

\maketitle

%--------------------------------
% INTRODUCTION
%--------------------------------
\paragraph{Introduction}
Molecular dynamics is an excellent,
although computationally very demanding, tool
to accurately predict rates for chemical reactions
and other activated barrier crossing processes.
Alternative, and simpler, approaches can
account for the reaction mechanism and rates,
often relying on dimensional reduction.
Transition State Theory (TST) \cite{Haenggi90,Miller,truh96} 
is among the most popular, because it provides a very 
\textit{simple} answer to the two most relevant issues 
in rate theory: 
to predict whether a trajectory is reactive or not, 
and to provide a simple expression to accurately 
compute the corresponding rates.
For this reason, TST has been used in fields far from the 
original chemical reaction dynamics where it was born, 
such as  celestial mechanics \cite{Jaffe02}, 
atomic ionization~\cite{Jaffe99}, 
surface science~\cite{Miret-Artes12},
or condensed matter~\cite{Wanasundara14}.

The fundamental problem that TST has faced since its inception 
is the correct identification of an optimal dividing 
surface (DS) separating reactants from products that is 
crossed \textit{once and only once} by \textit{all reactive} trajectories.
Although this DS must obviously sit somewhere 
close to the top of the energetic barrier between reactants and products, 
its exact geometry is critical, 
because
trajectories recrossing it give rise to an overestimation 
of the true rate constant.
A popular alternative is the variational TST (VTST)
that identifies the DS location by minimizing the 
number of recrossings
(see~\cite{Garrett05a} for a review).
Fortunately, it has been recently shown that using 
sophisticated geometrical techniques \cite{Uzer02,Waalkens04a,Waalkens04c} 
the problem can be  solved exactly for gas phase reactions.
For a reaction that is driven by a noisy environment with ohmic 
friction it can be solved if the DS itself is made time dependent~\cite{Bartsch05b,Bartsch05c,Bartsch06a,Bartsch08,Craven14,Craven14a,Craven15}.
Anharmonicities of the energy barrier can be taken into account 
perturbatively~\cite{Melnikov93, Kawai07a,Revuelta12,Bartsch12}.

In this Letter, we make TST exact
also in the more realistic, and more complicated,
case of non-Markovian friction.
Indeed, we show how to define
a rigorously recrossing-free DS in phase space.
This DS is time--dependent and moves randomly, 
``jiggling'' in the vicinity of the barrier.
By allowing a time-dependent DS,
we overcome the limits of fixed configuration space surfaces,
which often cannot be made recrossing-free,
as Mullen \textit{et~al.}~\cite{Gotchy14} have recently shown in several examples.

Even though the time-dependent DS satisfies the no-recrossing requirement of traditional TST,
a major advance can still be achieved by shifting the
focus away from the DS, which has to be arbitrarily
selected by hand, 
and onto invariant dynamical
structures that the system presents to us.
Specifically, we obtain a hypersurface in phase space that
unambiguously separates reactive from nonreactive trajectories.
In this way, reactive trajectories can be identified simply from their
initial conditions, without any laborious numerical simulation.
This separatrix,
which will be shown to be a stable manifold (SM),
provides both a more solid foundation and a more convenient practical tool
for rate theory than the conventional DS.
We compute the SM perturbativelly and thus obtain
an analytical expression for the transmission
factor and the rate constants for the crossing of
anharmonic potential barriers under non-Markovian noise.
We demonstrate the efficiency of our theory by
recoverring the correct reaction rates for a realistic 
model of the LiCN$\rightleftharpoons$LiNC isomerization
in an argon bath.

\begin{figure}
  \includegraphics[width=0.8\columnwidth]{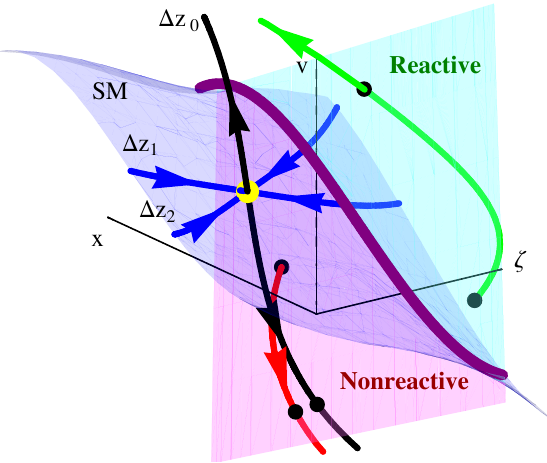}
\caption{
Geometric objects in phase space for Eqns.~\eqref{eq.Exp} 
with an anharmonic barrier.
The TS~trajectory is indicated by a yellow dot.
Its stable manifold (SM, light blue surface and trajectories therein)
and its unstable manifold (black curve) move and get deformed
randomly.
The purple curve  marks
the intersection of the SM with the  surface of initial conditions (plane $x=0$).
It partitions the surface of initial conditions into 
reactive (green) and nonreactive (red) regions
and defines the critical velocity~$V^\ddag(\zeta)$.
Representative reactive (green) and nonreactive (red) 
trajectories intersect the surface of initial conditions
as indicated by black dots.
}
\label{fig.1}
\end{figure}

Our current results shed new light on the surprising agreement
between PGH~theory~\cite{Pollak89}
and our earlier 
results~\cite{Muller08}
on the LiCN reaction at temperatures far above
the activated regime for which PGH theory was initially developed.
These results led 
Pollak and Ankerhold~\cite{Pollak13}
to revisit the assumptions of PGH theory.
They found that the bath temperature does not severely affect the energy loss terms
and hence does not modify the form of the rates.
In this Letter we obtain reaction rates in agreement with numerical simulations
from a different theoretical starting point,
and thus provide further confirmation that a rate description of the 
process is indeed appropriate.
Likewise, our results improve those reported by Pollak {\it et al.}~\cite{Pollak93, Pollak15},
where similar corrections to PGH
were obtained by applying a VTST to a Hamiltonian system whose dynamics
mimics that of the popular generalized Langevin equation (GLE)~\cite{Zwanzig01}, 
providing at the same time
a simpler and clearer picture of  the reaction mechanism
from a geometrical point of view (cf.~Fig.~\ref{fig.1}).

%--------------------------------
% METHOD
%--------------------------------

%
%%%%%%%%%%%%%%%%%%%%%%%%%%%%%%%%%%%%%%%%%
%Figure 1
%%%%%%%%%%%%%%%%%%%%%%%%%%%%%%%%%%%%%%%%%
\paragraph{Method}
For the sake of a simple presentation we restrict ourselves to systems 
with one degree of freedom (dof), although the generalization 
to higher dimensions is straightforward.
It will be reported elsewhere~\cite{Bartsch15}.

The reduced dynamics of a 1-dof system coupled to an external 
heat bath with memory effects can be adequately described 
by the GLE~\cite{Zwanzig01}
%
%Equation 1
\begin{equation}
	\label{genLE}
	m \ddot x = - \frac{d U(x)}{d x} 
             - m \int_{-\infty}^t \gamma(t-s)\, \dot x(s)\,ds 
             +m R_\alpha(t),
\end{equation}
where $m$ is the mass of the particle, 
$x$ its position, 
$U$ the potential of mean force, 
$\gamma(t)$ the friction kernel, and $R_{\alpha}(t)$ the fluctuating 
colored
noise force per unit mass exerted by the heat bath. 
It is related to~$\gamma(t)$ by the fluctuation-dissipation 
theorem, $\avg{R_\alpha(0) R_\alpha(t)}_\alpha = \kT\,\gamma (t)/m$,
where $\avg{...}_\alpha$ denotes the average over the different realizations
$\alpha$ of the noise.

If the friction kernel takes the exponential form
%
%Equation 2
\begin{equation}
	\gamma(t) = \frac{\gamma_0}{\tau}\, \exp (-t/\tau)
	\label{gammaExp}
\end{equation} 
with a characteristic correlation time~$\tau$ and 
a damping strengh~$\gamma_0$,
%\CB{a damping strengh~$\gamma_0=\int_0^\infty \gamma(t) dt$},
the GLE~\eqref{genLE} can be replaced by a system of differential 
equations on a finite dimensional extended phase 
space~\cite{Ferrario79, Grigolini82, Marchesoni83,Martens02}
%
%Equation 3
%\begin{align}
%	\label{eq.Exp}
%	\dot x & = v, \nonumber \\
%	\dot v & = \omb^2x + f(x) + \zeta,  \\
%	\dot \zeta & = -\frac{\gamma}{\tau}\,v - \frac{1}{\tau}\,\zeta 
%                   + \xi_\alpha(t), \nonumber
%\end{align}
\begin{equation}
	\label{eq.Exp}
	\dot x = v, \;
	\dot v = \omb^2x + f(x) + \zeta,  \;
	\dot \zeta = -\frac{\gamma_0}{\tau}\,v - \frac{1}{\tau}\,\zeta 
                   + \xi_\alpha(t), 
\end{equation}
where the mean force $-d U(x)/d x = m \omb^2x + m f(x)$ 
is split into a linear term and non-linear corrections
$f(x)=-\varepsilon c_3 x^2 - \varepsilon^2 c_4 x^3 - \dots$.
The perturbation parameter $\varepsilon$ measures the 
anharmonicity of the barrier potential and  will be set equal to 1 
at the end of the calculation.
The auxiliary coordinate~$\zeta$ is given by
$\zeta = -\int_{-\infty}^t \gamma(t-s)\,\dot x(s)\,ds$,
and~$\xi_\alpha$ is a white noise source 
satisfying the fluctuation--dissipation theorem
$\avg{\xi_\alpha(t)\;\xi_\alpha(s)}_\alpha=[2\kT\,\gamma_0/
(m \tau^2)]\delta(t-s)$.

If $f(x)=0$, the equations of motion~\eqref{eq.Exp} are linear
and can be solved by diagonalizing the coefficient matrix.
We find one positive eigenvalue~$\lambda_0$ and two eigenvalues~$\lambda_{1,2}$
that are negative or have negative real parts.
The corresponding diagonal coordinates are denoted by~$z_i$.

Equations~\eqref{eq.Exp} have a unique solution, called the 
TS~trajectory~\cite{Bartsch05b, Bartsch05c, Kawai07a, Revuelta12,Bartsch12}  
that remains ``jiggling'' in the vicinity 
of the saddle point for all times.
It depends on the realization~$\alpha$ of the noise.
We denote its diagonal coordinates by~$z_i^\ddag(t)$ and
its position by~$x^\ddag(t)$.
For the harmonic barrier, i.e.~$f(x)=0$, the coordinates~$z_i^\ddag(t)$
can be obtained explicitly as an integral over the noise~$\xi_\alpha$~\cite{Bartsch05c, Revuelta12,Bartsch12}.
The TS~trajectory gives rise to a time-dependent DS $x=x^\ddag(t)$
that is recrossing-free in the harmonic 
approximation~\cite{Bartsch05b,Bartsch05c} 
as well as in anharmonic systems~\cite{Craven14,Craven14a,Craven15}.
However, we will not consider this~DS any further
and focus instead on the invariant structures that
determine the reaction dynamics.

In relative coordinates 
$ 
    \Delta z_i = z_i-z_i^\ddag, 
$ Eq.~\eqref{eq.Exp} reads
%
%Equation 4
\begin{equation}
        \label{eq.dDzidt}
	\Delta \dot z_i = \lambda_i \, \Delta z_i +K_i\,f(x).
\end{equation}
Here~$K_i=-(\lambda_j+\lambda_k)/[(\lambda_i-\lambda_j)(\lambda_i-\lambda_k)]$, where $i,j,k$ take the values $0,1,2$ and must be different.
In the harmonic limit Eq.~\eqref{eq.dDzidt} has the simple solution
$\Delta z_i(t)=\Delta z_i(0) \exp(\lambda_i t)$.
Thus, as $\lambda_0>0$, $\Delta z_0(t)$ is associated with an exponentially 
growing unstable direction in phase space,
whereas~$\Delta z_1(t)$ and~$\Delta z_2(t)$
are both associated with stable directions.
The plane $\Delta z_0=0$ forms the SM of the TS~trajectory.
Trajectories within it asymptotically approach the TS~trajectory as $t\to\infty$;
they are trapped near the barrier top.
Because the SM contains trajectories that are neither reactive nor nonreactive,
it separates reactive from nonreactive trajectories.

When anharmonic terms are present, 
the SM is deformed in a time-dependent manner, 
but it stills remains the separatrix between reactive and nonreactive 
trajectories:
All trajectories starting above the SM 
approximate the unstable manifold for large positive values
of $\Delta z_0$ and finish in the product well defined by~$x>0$, 
while trajectories that lie below the SM will follow the negative 
part of the unstable manifold into the reactant well $x<0$, 
as sketched in Fig.~\ref{fig.1}.

\paragraph{Reaction rates}
The reaction rate can be computed by sampling trajectories from a 
Boltzmann ensemble at the barrier top and calculating the reactive 
flux across the surface of initial conditions $x=0$.
Under the TST assumption that this surface is recrossing free,
i.e. a trajectory is reactive if it starts with an initial velocity $v>0$,
this procedure yields a reaction rate $k^\text{TST}$ that overestimates 
the true rate~$k^{\rm exact}$.
The violation of the TST~assumption can be quantified by the transmission 
factor~$\kappa=k^{\rm exact}/k^\text{TST} \le 1$.
The exact rate is obtained if the flux calculation includes only 
trajectories that are actually reactive.
These are the trajectories that lie above the SM,
or, as Fig.~\ref{fig.1} shows, whose initial velocity is larger 
than a critical velocity $V^\ddag(\zeta)$ that depends on the realization 
of the noise and on the initial value of the auxiliary coordinate~$\zeta$.
This critical velocity encodes 
all the relevant information about the reaction dynamics.
Because it leads to an exact characterization of reactive trajectories,
the critical velocity and the SM that determines it
are more fundamental to the theory
than the DS that has customarily been used.
We compute the critical velocity by a perturbative expansion~$V^\ddag
 = V^{\ddag(0)} + \varepsilon V^{\ddag(1)}  + \varepsilon^2 V^{\ddag(2)} + \dots$.
This computation follows the method developed in Refs.~\onlinecite{Revuelta12,Bartsch12} 
for the case of Markovian friction.
Full details will be presented elsewhere \cite{Bartsch15}.

Equipped with the critical velocity 
one can compute~\cite{Bartsch08, Revuelta12, Bartsch12} the
transmission factor 
$\kappa = \avg{e^{-V^{\ddag\,2}/2\kT}}_{\alpha,\zeta}$,
which is averaged both over the noise~$\alpha$ and 
the initial value of~$\zeta$. 
Now, by expanding~$\kappa$ 
as~$\kappa=\kappa_0 +\varepsilon \kappa_1 +\varepsilon^2 \kappa_2+\ldots$, 
we finally obtain its lowest order 
%
%Equation 5
\begin{equation*}
   \kappa_0  =  \frac{\lambda_0}{\omb}, \qquad \kappa_1 =  0, \label{eq.kappa0}
\end{equation*}
\begin{widetext}
\begin{align}
   \kappa_2 & = -\frac{3 \kappa_0 \kT}{4 m \omb^4}
                       \left( \frac{ f^{0,0}_{0,1,-1} }{f^{0,0}_{1, \eta-1, \eta} } \right)^2
                      \! \! \! \bigg\{
	\frac{2 c_3^2  \left[
	 	f^{2,4}_{110, 329, -12}+ 
		5 f^{10, 0}_{4, -17, 4} +
                2 \left( f^{0,5}_{10, 41, 10}+
	                   f^{4,3}_{115, 197, -28} +
                           f^{6, 2}_{115, 22, 8} + 
		           f^{8, 1}_{55, -94, 6} \right) \right]}
	{9 \, \omb^2  \, 
               f^{0,0}_{0, 1,  \eta} \, 
               f^{0,0}_{1, 2 \left(\eta-2\right), 4 \eta}
               f^{0,0}_{4,2 \eta-1, \eta}} 
        + c_4 f^{0,0}_{0,1,\eta} \bigg\}, \label{eq.kappa2}
\end{align}
with~$\eta = \lambda_0(1+\lambda_0\tau)/(\omb^2\,\tau)$, and~$f^{a,b}_{c,d,e}= 
\kappa_0^a \, \eta^b \left( c \,  \kappa_0^4 + d \, \kappa_0^2 + e \, \right) $.
\end{widetext}
%\CB{with~$\eta = \lambda_0(1+\lambda_0\tau)/(\omb^2\,\tau)$, and~$f^{a,b}_{c,d,e}= 
%\kappa_0^a \, \eta^b \left( c \,  \kappa_0^4 + d \, \kappa_0^2 + e \, \right) $.}
The leading order~$\kappa_0$ recovers the 
well known  Grote--Hynes theory (GHT)~\cite{Grote80}.
Because all odd order terms are zero,
the perturbation expansion proceeds in powers of~$\kT$.

\paragraph{Model} 
To illustrate the performance of our method we apply it to a simple,
yet realistic, model for the LiNC$\rightleftharpoons$LiCN isomerization.
It has a number of properties that make it very attractive for
dynamical studies.
Most importantly, % the most important being that
the bending mode in this system is very floppy, 
so that chaos sets in at moderate values of the excitation energy. 
This reaction has been extensively studied by some of us in the past
and very recently in connection to THz reactivity 
control~\cite{PellouchoudReed15}.
Most relevant in the present context,
it furnished the first observation \cite{Muller08,Muller12} of the turnover predicted 
by Kramers in his 1940 seminal paper~\cite{Kramers40,Pollak89,Pollak13}.

\begin{figure}
\includegraphics[width=0.8\columnwidth]{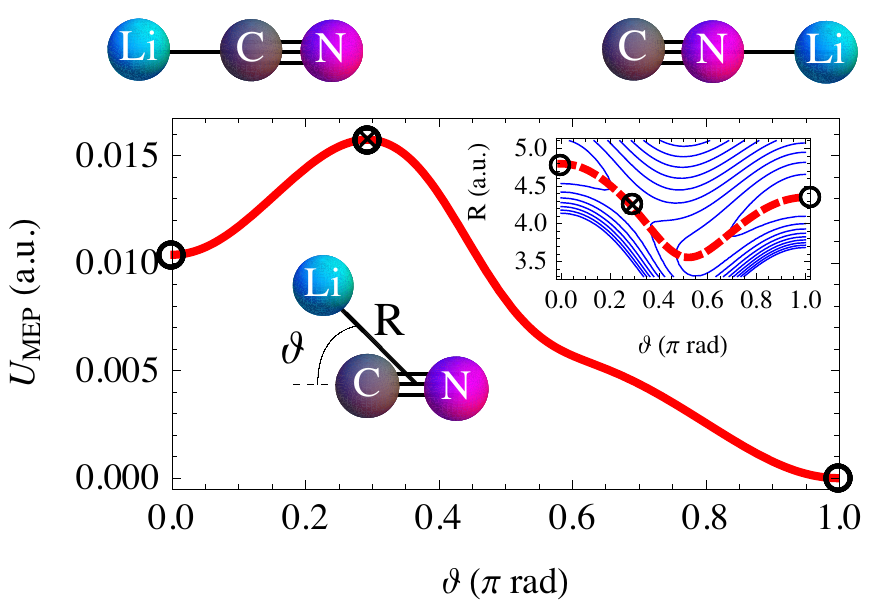}
\caption{Effective potential for the one-dimensional model of LiCN isomerization.
It corresponds to the minimum energy path connecting 
the two potential wells of the LiNC/LiCN molecular system.
The configurations at the 
barrier top 
(crossed circle),
and of the two stable isomers 
associated with the well minima 
(open circles) are also shown.
Inset: Contour plot of the 2-dof  potential.
The minimum energy path is plotted superimposed in dashed red line.}
\label{fig.2}
\end{figure}

To describe the configuration of the LiCN molecule,
we use the distance $r$ between the C and N atoms,
the distance $R$ of the Li atom from the center of mass of the CN fragment
and the angle $\vartheta$ between the Li atom and the CN axis
(see Fig.~\ref{fig.2}).
Because the CN triple bond is very rigid,
the distance $r$ will not deviate much from its equilibrium value 
$r_e=2.186\ \text{a.u.}$
A potential energy function describing the motion of the Li atom relative  
to a rigid CN was introduced by Essers \textit{et al.} \cite{Essers}.
An improved model can be obtained by combining this potential with a 
Morse potential for the CN vibration~\cite{Muller3D}.
The potential energy of the molecule with $r=r_e$ is shown in 
the inset to Fig.~\ref{fig.2}.
It has two wells at $\vartheta=0$ and $\vartheta=\pi$~rad
that correspond to the two  linear isomers Li--CN and Li--NC. 

Extensive molecular dynamics (MD) simulations of this molecule in a 
bath of 512~argon atoms were reported in Refs~\cite{Muller08,Muller12}.
It was found there that the isomerization rates for the transitions 
from the Li--NC to Li--CN configuration and back can be well described 
by a one-dimensional model in which the molecule is assumed to move 
along the minimum energy path (MEP).
The MEP and the corresponding potential energy profile are shown in Fig.~\ref{fig.2}.
This effective potential yields the parameters in Table~\ref{tab.I} that will be 
used in perturbation theory.

In our study, the dynamics is modeled by the GLE~\eqref{genLE}, 
in which the angle~$\vartheta$ plays the role of the position~$x$
and the potential $U$ is the MEP potential $U_\text{MEP}$ of Fig.~\ref{fig.2}.
The mass~$m$ is replaced by the moment of inertia $I_\vartheta$ that 
describes the rotation of the Li atom relative to the CN~fragment.
Though the value of $I_\vartheta$ varies along the MEP,
in the spirit of TST it is fixed to its value at the saddle point 
of the potential,  $I_\vartheta=42\,852\ \text{a.u.}$
The friction kernel is well approximated by the exponential form~\eqref{gammaExp}
with the decay time $\tau=0.84 \gamma_0/\omega_b^2$~\cite{Muller3D}.
%
%%%%%%%%%%%%%%%%%%%%%%%%%%%%%%%%%%%%%%%%%
%Table I
%%%%%%%%%%%%%%%%%%%%%%%%%%%%%%%%%%%%%%%%%
\begin{table}
  \caption{Parameters of the effective potential shown in
Fig.~\ref{fig.2} for the two well minima (isomers) 
and the saddle point.}
  \label{tab.I}
  \begin{tabular}{cccc}
   \hline
Parameter & Li--CN  & Saddle point & Li--NC \\
   \hline
   \hline
$\vartheta$ (rad)                & 0      &  0.917 & $\pi$ \\
$U_\text{MEP}$ ($10^{-2}$ a.u.)              & 1.04   & 1.58   & 0     \\
    $\omega$ ($10^{-4}$ a.u.)     & 7.92   & 9.65   & 5.90  \\
    $c_3$ ($10^{-7}$ a.u.)        & --     & -8.0   & --    \\
    $c_4$ ($10^{-7}$ a.u.)        & --     & 7.4    & --    \\
   \hline
  \end{tabular}
\end{table}
%
%%%%%%%%%%%%%%%%%%%%%%%%%%%%%%%%%%%%%%%%%%
%Figure 2
%%%%%%%%%%%%%%%%%%%%%%%%%%%%%%%%%%%%%%%%%%

\begin{figure}
\includegraphics[width=0.8\columnwidth]{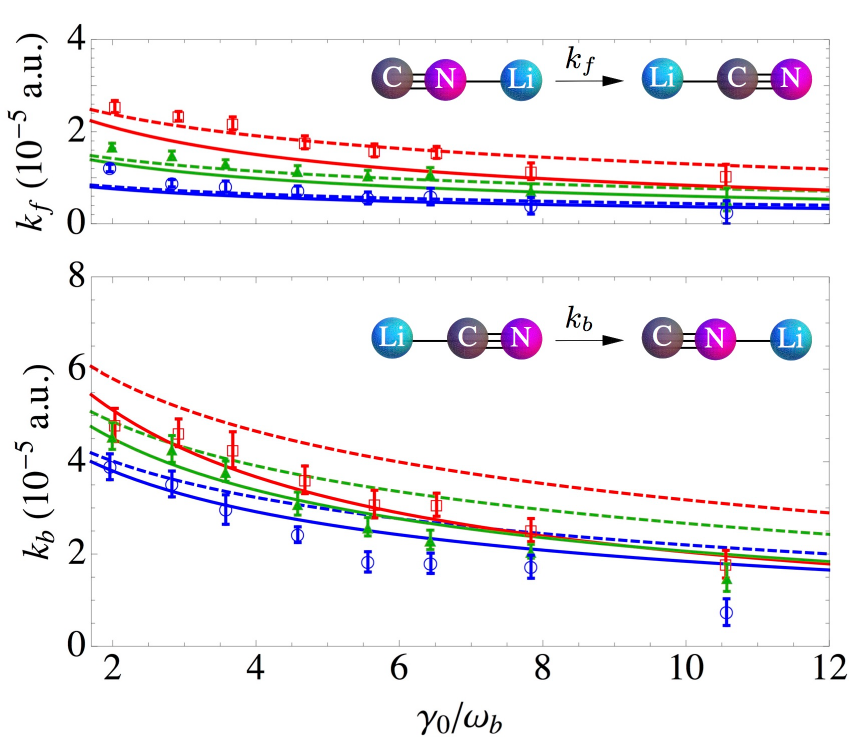}
\caption{Reaction rates for the forward (top) and backward (bottom) 
 LiNC$\rightleftharpoons$LiCN isomerization
 as a function of the bath friction.
 Perturbation theory results of order zero (dashed) and order two (solid) are shown
 for temperatures $T$=2500K (blue), 3500K (green), and 5500K (red).
 For comparison, results of all-atom MD simulations are shown by symbols with one-sigma error bars.
}
\label{fig.3}
\end{figure}

%%%%%%%%%%%%%%%%%%%%%%%%%%%%%%%%%%%%%%%%%%
\paragraph{Results}
In Fig.~\ref{fig.3}, our predictions from perturbation theory (PT)
for both the forward LiNC$\rightarrow$LiCN (top), and 
backward LiCN$\rightarrow$LiNC (bottom) reactions as a function 
of the adimensional friction $\gamma_0/\omb$
are compared with the results of all-atom MD simulations.
Results are presented for temperatures
$T$=2500K (blue), 3500K (green), and 5500K (red).
Perturbative results in orders~0 and~2 are indicated by 
dashed and full lines, respectively.
Because our rate theory, like GHT, is only valid in the spatial diffusion limit, 
where the friction has moderate to strong values, results for 
$\gamma_0/\omega_b < 2$ are not included in Fig.~\ref{fig.3}.
From the comparison, the following comments can be made.

The rates always increase with temperature,
as should be expected for an activated process.
The rates of the forward reaction  are smaller than those of the
backward reaction since the corresponding energy barrier is larger.
The perturbative correction is negative.
Its magnitude increases with temperature,
as expected from Eq.~\eqref{eq.kappa2}.
For the backward reaction,
where the second-order correction is large,
it provides a clear improvement of GHT for all values of the parameters.
For the forward reaction, the second-order correction is barely
noticeable at low temperatures.
At the highest temperature $T=5500$K,
where the perturbative correction is significant,
the MD results are closer to GHT than to the PT results if damping is weak.
For high damping, the second-order PT again provides a marked improvement 
over GHT.

In all cases, there is excellent agreement between the MD and PT rates.
In fact, the agreement is striking, considering that the MD results
were obtained from a simulation with an explicit argon bath
that is much more complex than the simple one-dimensional model
that yields the PT results.

%
%%%%%%%%%%%%%%%%%%%%%%%%%%%%%%%%%%%%%%%%%
%Figure 3
%%%%%%%%%%%%%%%%%%%%%%%%%%%%%%%%%%%%%%%%%

%--------------------------------
% CONCLUDING REMARKS
%--------------------------------
%\newpage
\paragraph{Concluding remarks}
In summary, it is possible in principle
to define a time-dependent recrossing--free DS in phase space 
for the dynamics of a particle in an anharmonic barrier that 
interacts with the environment via non--Markovian friction,
i.~e. via colored noise force.
However, we have demonstrated that it is advantageous
to base a rate calculation on invariant geometric structures,
namely the SM of the TS~trajectory,
instead of a DS, as  customary in TST.
The SM allows the unambiguous identification of 
reactive trajectories simply by inspection of their initial conditions,
without having to resort to any time--consuming numerical 
simulation.
It provides a formally exact rate formula
that we have evaluated through perturbation theory.
In this way we have obtained 
an explicit expression for the transmission factor
that corrects GHT by including 
anharmonic effects.
It agrees well with the results of an all-atom model 
of LiCN isomerization in an argon bath.
Finally, the method outlined here
can be straightforwardly generalized to systems 
of higher dimensionality, 
as will be reported elsewhere~\cite{Bartsch15}.

%--------------------------------
% ACKNOWLEDGMENTS
%--------------------------------
\paragraph{Acknowledgments}
We gratefully acknowledge support from the
Minis\-terio de Econom\'ia y Competitividad (Spain) 
under Contracts No.\ MTM2012-39101 and MTM2015-63914-P,
and ICMAT Severo Ochoa SEV-2011-0087 and SEV-2015-0554. 
Travel between partners was partially supported
through the People Programme (Marie Curie Actions)
of the European Union's Seventh Framework Programme
FP7/2007-2013/ under REA Grant Agreement No. 294974.

\bibliography{ColouredRateLetter}

\end{document}